\documentclass[aps,prb,twocolumn,showpacs,amsmath,amssymb,superscriptaddress]{revtex4-1}
\usepackage{graphicx}
\usepackage{dcolumn}
\usepackage{epstopdf}
\usepackage{bm}
\usepackage{color}

\begin{document}
\title{Graphene adhesion on mica: Role of surface morphology}

\author{A.~N. Rudenko}
\email[]{rudenko@tu-harburg.de}
\author{F.~J. Keil}
\affiliation{Institute of Chemical Reaction Engineering, Hamburg University of Technology, Eissendorfer Strasse 38, D-21073 Hamburg, Germany}
\author{M.~I. Katsnelson}
\affiliation{Institute for Molecules and Materials, Radboud University Nijmegen, Heijendaalseweg 135, 6525 AJ Nijmegen, The Netherlands}
\author{A.~I. Lichtenstein}
\affiliation{Institute of Theoretical Physics, University of Hamburg, Jungiusstrasse 9, D-20355 Hamburg, Germany}
\date{\today}

\begin{abstract}
We investigate theoretically the adhesion and electronic
properties of graphene on a muscovite mica surface using the density
functional theory (DFT) with van der Waals (vdW) interactions
taken into account (the vdW-DF approach). We found that
irregularities in the local structure of cleaved mica surface
provide different mechanisms for the mica-graphene binding.
By assuming electroneutrality for both surfaces, the binding is mainly
of vdW nature, barely exceeding thermal energy per carbon atom at
room temperature. In contrast, if potassium atoms are
non uniformly distributed on mica, the different regions of the
surface give rise to $n$- or $p$-type doping of graphene. In turn,
an additional interaction arises between the surfaces,
significantly increasing the adhesion. For each case the
electronic states of graphene remain unaltered by the
adhesion. It is expected, however, that the Fermi level of graphene 
supported on realistic mica could be shifted relative to the Dirac point 
due to asymmetry in the charge doping. Obtained variations of the 
distance between graphene and
mica for different regions of the surface are found to be
consistent with recent atomic force microscopy
experiments. A relative flatness of mica and the absence of
interlayer covalent bonding in the mica-graphene system make 
this pair a promising candidate for practical use.
\end{abstract}

\maketitle

\section{Introduction}
A monolayer of graphite, commonly known as graphene, is the first
truly two-dimensional crystal (one atom thick), which became
experimentally available within the last few
years.\cite{Novoselov,APSnews} Remarkable electronic properties of
graphene make this material a promising candidate for a large
variety of electronic
applications.\cite{Geim,Graphene-RMP,Geim_Science}

Usually graphene is deposited on different substrates owing to the
peculiarities of preparation techniques.\cite{mm_cleavage,epitaxy}
The role of substrates and their effect on electronic
transport in graphene are actively debated, but are still not clearly
understood. Meanwhile, a number of experimental and theoretical
studies show that many properties of graphene are strongly
dependent on the substrate.\cite{Zhou,Wehling-APL,Wintterlin}

Being a two-dimensional crystal, the free standing graphene is not
atomically flat but possesses intrinsic corrugations of the
structure (ripples), due to thermal bending
fluctuations.\cite{Meyer,fasolino} Although scanning tunneling
microscopy (STM) and atomic force microscopy (AFM) experiments
allow one to reveal corrugations of graphene on insulating surfaces
(e.g., SiO$_2$),\cite{ishigami,geringer} the existence of
intrinsic ripples in substrate-supported graphene is still a
subject of discussion. Being supported on a surface, the
corrugated graphene structure may simply reflect the conformation
between graphene and the underlying substrate. Quite recently it
was reported that the intensity of such ripples can be strongly
dependent on the substrate on which graphene is
deposited.\cite{chun} In particular, graphene monolayers display
an exceedingly flat structure being placed on a mica surface, which
is several times smoother than a SiO$_2$ surface. This observation
means that the ripples, independently on their nature, can be
strongly suppressed by interfacial interactions between graphene
and an appropriately chosen (flat) substrate. General theoretical
models also support this suggestion.\cite{li-zhang}

As for the substrate, micas are known to be well suited for
fundamental studies as well as for technological purposes owing to
their relative atomic smoothness and a large band gap (7.85
eV).\cite{davidson} These properties make this material a
favorable candidate as a substrate for the deposition of graphene
in potential graphene-based devices. Although experimental studies
propose a strong interfacial binding between graphene and mica
resulting from the vdW interaction,\cite{chun,lee} the nature of
such a binding has not been unambiguously established. Details of
the binding, such as dependence on the surface morphology, are also
unclear.

In this work we examine the adhesion and electronic properties of
graphene supported on a muscovite mica surface by means of
first-principles methods. By assuming certain atomic disorder of the mica
surface we found that binding with graphene can vary significantly
from one surface region to another due to the charge-transfer
doping of graphene. As a result, graphene might adopt its lattice
accordingly, giving rise to a wavy like structure, but such
corrugations of the graphene structure turned out to be rather
small. An estimation of maximum height variation shows
reasonable agreement with topographic data of AFM.\cite{chun} 
We show that the typical
electronic structure of graphene remains unperturbed being in
contact with the mica surface, which plays an important role in
practical applications of graphene. The possible influence of 
the mica substrate on transport properties of graphene is also addressed.  

The paper is organized as follows. In Secs.~II~A and II~B we briefly
describe the computation methods and crystal structures of
the investigated systems, respectively. Section III is devoted to the
results and their analysis. In Sec.~IV we summarize our results.

\section{Computational details}

\subsection{Calculation method}

Ground-state energies and electronic density distributions have
been calculated using the plane-wave pseudopotential method as
implemented in the {\sc quantum-espresso} simulation
package.\cite{espresso,pseudo} In order to calculate adsorption
energies and properly take into account dispersive interactions,
we use the vdW-DF approach proposed by Dion \emph{et
al.}\cite{Dion,Thonhauser} showing transferability across a broad
spectrum of interactions.\cite{Langreth,CAD} In this method, the
exchange-correlation energy functional consists of three parts:
(i) the exchange part of the revised Pedrew-Burke-Ernzerhof (revPBE)
functional,\cite{Zhang-Yang} (ii) the local correlation part of
the standard local density approximation (LDA) functional, and (iii)
the non local correlation part, incorporating effective many-body
density response and allowing treatment of dispersive interactions
without any fitting parameters.

In our calculations we employed an energy cutoff of 30 Ry for the
plane-wave basis and 300 Ry for the charge density.
Self-consistent calculations of the Kohn-Sham equations were
carried out imposing the convergence criterion of 10$^{-8}$ Ry.
For Brillouin-zone integration, the tetrahedron
scheme\cite{tetrahedron} and (16 $\times$ 8 $\times$ 1)
Monkhorst-Pack {\bf k}-point mesh \cite{monkhorst} were used.
A much finer mesh (48 $\times$ 24 $\times$ 1) and a Gaussian broadening of
0.02 Ry were used for the density of states (DOS) calculations.

In order to find the ground structural states of the investigated
systems we performed a relaxation of the supercell with fixed
in-plane lattice parameters. The stop criterion for the relaxation
was set to 0.001 Ry/\AA, except for the lowermost layer of atoms,
whose positions had been fixed. For all the cases under
consideration the height of the supercell was chosen to be 50 \AA.
In order to avoid spurious interaction between images of the
supercell in the [001] direction we also used a dipole
correction.\cite{dc}

\subsection{Surface structures}

Graphene has a two-dimensional honeycomb lattice of $sp^2$-bonded
carbon atoms. Although the real structure of graphene is
corrugated, the characteristic length of corresponding ripples is
around 100 \AA,\cite{Meyer,fasolino} which is much
larger than the typical length of the supercell used in
first-principles calculations. For this reason we do not consider
this phenomenon in our work directly.

We take into account distortions of carbon lattice caused by
non uniformity of the substrate, though these distortions are
found to be negligibly small due to the strong $sp^2$ bonding
between carbon atoms and imposed boundary conditions. In general,
graphene lattice is not commensurate with the substrate in lateral
directions. To overcome this issue we slightly adjust the lateral unit
cell parameters of the substrate as described below, bearing in mind 
that micas have a relative low bulk modulus, i.e., they can be compressed quite
easily.\cite{min_handbook} We used the lattice constant of
graphene equal to $a=2.459$ \AA \ in accordance with the
experimentally obtained value for graphite at low
temperatures.\cite{Baskin}

\begin{figure}[!tbp]
\includegraphics[width=0.40\textwidth, angle=0]{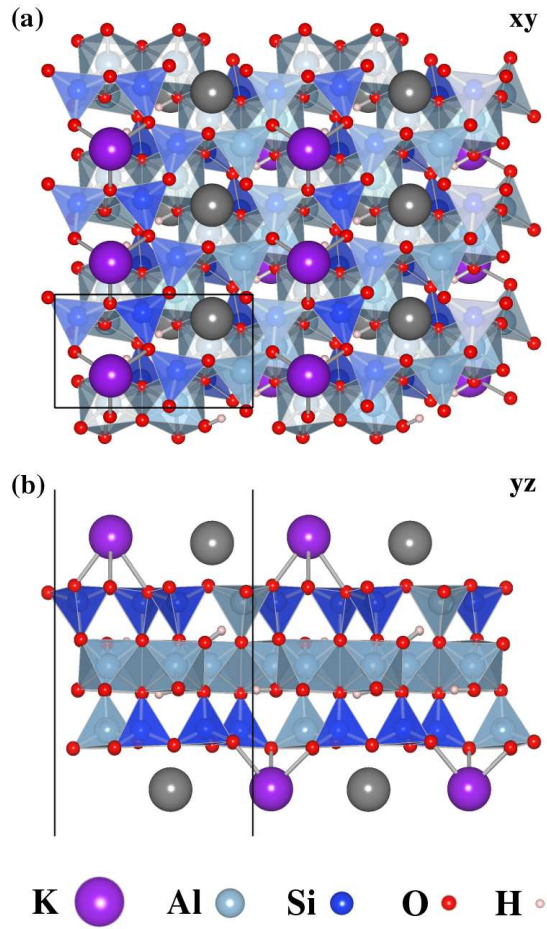}
\caption{(Color online) Surface structure of muscovite mica:
(a) [001] projection and (b) [100] projection. Depicted structure corresponds to
the electroneutral surface with uniform distribution of K$^{+}$ ions. Disconnected
gray spheres show the positions of extra K$^{+}$ ions in the case of
an electropositive surface.
Black solid lines are the boundaries of the unit cell.}
\label{mica_struct}
\end{figure}

Micas belong to the group of phyllosilicate minerals exhibiting a
two-dimensional sheet structure. In this work we examine the
surface of muscovite, the most abundant variety of mica. Muscovite
is a 2:1 layered dioctahedral aluminosilicate with the formula
KAl$_2$(Si$_3$,Al)O$_{10}$(OH)$_2$.\cite{mica_cryst} Structurally,
each irreducible muscovite layer consists of one layer of
octahedrally coordinated Al$^{3+}$ ions, which is sandwiched
between two tetrahedral silicate layers with vertices pointing
toward the octahedral layer (Fig. \ref{mica_struct}). Within
tetrahedral units aluminum is randomly substituted for silicon
with a ratio of 1:3. To compensate the negative charge of adjacent
mica layers, potassium counterions are present in 12-fold oxygen
coordination.

After the cleavage, half of the potassium ions are assumed to be
left to preserve electroneutrality of the surface as a whole.
However, the positions of the ions and their distribution over the
surface are not well defined from the experimental point of view.
Since the interaction between potassium and the surface is of
ionic nature, the binding is strong enough to prevent diffusion of
potassium ions across the surface at room temperatures. Because
lateral diffusion is not possible, a uniform distribution of
K$^+$ cannot be established and, therefore, various regions of the
surface could be electrically charged.

Although the ionic surfaces are very reactive and easily adsorb
impurities from the environment to neutralize themselves, the
existence of uncompensated charges on the mica surface is
experimentally verified. Previously, it has been found that the
surface potential as well as the surface charge of freshly cleaved
mica are sensitively dependent on the environment
composition.\cite{qi} Moreover, the surface potential of mica
cleaved under ultra high vacuum (UHV) conditions is up to two
orders of magnitude higher than that for mica cleaved in
air.\cite{ostendorf} Therefore, we suppose that the strong
charging of UHV-cleaved mica is associated with the presence of
non uniformly distributed K$^{+}$ ions over the mica surface.

To take into account surface-charge effects, we consider the
following possibilities for the surface structure: (i)
electroneutral structure with uniform distribution of K$^{+}$ ions; 
(ii) electropositive structure with
double K$^{+}$ coverage; and (iii) electronegative structure in the
absence of K$^{+}$ ions. We note that for all the cases the whole
supercell remains neutral. The change of the surface type is achieved
only by varying the concentration of K adatoms and \emph{not} by varying
the number of valence electrons in the system.

In the case of the electroneutral substrate,
the supercell used in our study consists of a 42-atomic slab of
mica and a 16-atomic graphene layer. In order to match graphene
and mica supercells, we use a slightly compressed unit cell of
mica and employ the following in-plane parameters: $(2 \times
2\sqrt 3)a$, where $a$ is the lattice constant of graphene. This
choice corresponds to an $\sim$6\% decrease of the mica
experimental lattice constant. The volume of the bulk mica unit
cell with given in-plane parameters is about the same as predicted by
the LDA.\cite{ortega-castro}
Initial atomic positions for mica
were taken from neutron diffraction data,\cite{rothbauer} and
subsequently relaxed.

It should be noted that there are a number of ways to deposit graphene
on the mica surface. In our study we employ the configuration in which
lateral coordinates of the topmost potassium atoms on mica are 
maximally close to that of the center of the carbon hexagon. This choice
seems to be reasonable since it corresponds to the most stable 
configuration of single potassium ions adsorbed on graphene.\cite{Chan,jin}

\section{Results and discussion}

\begin{figure}[!tbp]
\includegraphics[width=0.40\textwidth, angle=0]{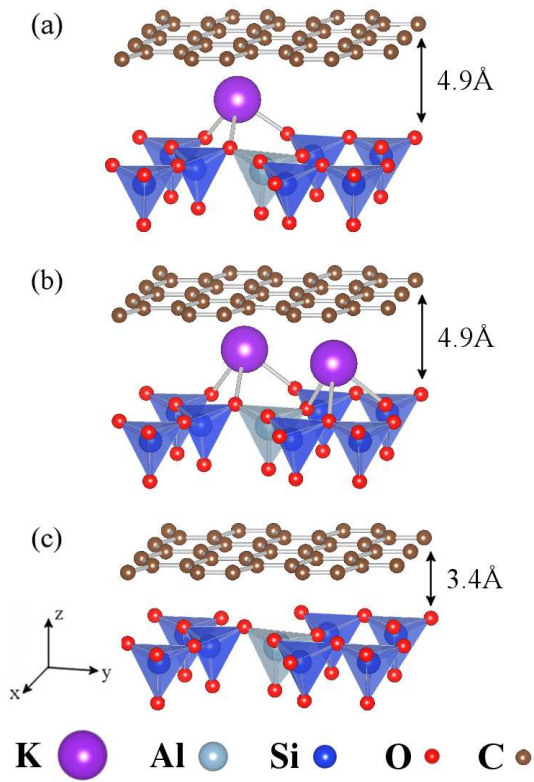}
\caption{(Color online) Equilibrium structure of graphene supported on a (a) neutral mica surface, (b) positive mica surface, and (c) negative mica surface. Only the topmost tetrahedral layer of mica is
shown.}
\label{mica_graphene}
\end{figure}

In Fig.\ref{mica_graphene} we show relaxed atomic structures of the
mica-graphene interface. One can see that there are no significant
changes in the surface structures of mica compared to the bare surfaces
(Fig.~\ref{mica_struct}). The only structural parameter affected by the
adhesion is the distance between graphene and the topmost oxygen layer
of mica. This distance is larger for electroneutral 
[Fig.~\ref{mica_graphene}(a)] and electropositive [Fig.~\ref{mica_graphene}(b)] 
cases due to the presence of K$^{+}$ ions on the surface.
(Figures \ref{mica_struct} and \ref{mica_graphene} 
were generated using the {\sc vesta} program.\cite{vesta})

We summarize adhesion energies and equilibrium interlayer distances for
graphene adsorbed on mica in Table \ref{mica_table}. In this table
we also show the part of the vdW interaction that contributed to the
total energy calculated as a difference between adhesion energies
in the presence of the non local correlation functional and without
it. As can be seen, the adhesion energies as well as the nature of
the interface interaction are strongly dependent on the surface type.
For the electroneutral mica surface the adhesion is caused primarily
by the vdW interaction. In this particular case the binding between
graphene and the mica surface is quite small in comparison with the
interlayer binding in graphite ($\sim$50 meV/C).\cite{halo} In the
case of both electropositive and electronegative mica surfaces the
contributions of the vdW interaction are much smaller than for the
neutral surface. Nevertheless, the adhesion energy in these two
cases is much stronger than for the neutral case, and exceeds the
interlayer binding in graphite. This indicates that there is
another mechanism of adhesion besides the vdW interaction.

    \begin{table}[!bt]
    \centering
    \caption[Bset]{Calculated adhesion energies and equilibrium interface distances for graphene supported on a mica surface. Results are given for three
different types of mica surfaces as discussed in Sec.~II~B. Adhesion energies are given in meV per carbon atom. $WF$ and $EA$ correspond to work functions and electron
affinities for the bare mica surfaces.}
    \label{mica_table}
\begin{ruledtabular}
 \begin{tabular}{ccccc}
                           & $e^{-}$-neutral & $e^{-}$-positive & $e^{-}$-negative   \\
     \hline
  $E_{adh}$, meV/C\footnotemark[1]         &  -29.3     & -75.5    &  -114.8     \\
 vdW part of $E_{adh}$     &   92\%   &  69\%  &   53\%      \\
  $d_{eq}$, \AA \footnotemark[2]           &   4.9      &  4.9     &   3.4       \\
  $WF$, eV            &   4.15      &  2.82     &   9.09       \\
  $EA$, eV            &   1.25      &  2.82     &   9.09       \\

      \hline
    \end{tabular}
\end{ruledtabular}
\footnotetext[1]{Adhesion energy is calculated in the standard way, i.e. as a difference between the total energies of the mica-graphene system and its isolated components.}
\footnotetext[2]{Interface distance between graphene and the mica surface implies the difference between averaged $z$ coordinates of carbon atoms in graphene and oxygen atoms in the
topmost tetrahedral layer of mica.}
    \end{table}

As has been demonstrated in previous studies, an impurity doping
of graphene as well as contact with substrates can lead to an
electronic transfer in the
system.\cite{halo,dai,imp_wehling,Khomyakov,sque,Chan} Let us
consider the possibility of this phenomenon in our case. For two
physical systems being in contact the charge transfer occurs if
the electronic affinity ($EA$) of one system is larger than the
work function ($WF$) of the other, if the process is
energetically favorable. In Table \ref{mica_table} we show $WF$s
and $EA$s for the bare mica surfaces considered, calculated as
($E_{\mathrm{vac}}-E_F$) and ($E_{\mathrm{vac}}-E_{\mathrm{cond}}$), where $E_{\mathrm{vac}}$, $E_F$,
and $E_{\mathrm{cond}}$ are the vacuum level of the electrostatic
potential, the Fermi energy, and the lowest energy of the
conduction band, respectively. Comparing the given values with the
graphene $WF$ (which turns out to be equal to $EA$ because of the
absence of a band gap) of $4.21$ eV, we see that one can expect
$n$-type doping for graphene supported on the electropositive mica
surface, $p$-type doping in the case of the electronegative surface,
and also tiny $n$-type doping in contact with the neutral
substrate since the surface $WF$ is almost equal to the graphene
$EA$ in this case. Furthermore, one can notice that the larger the
difference between the acceptor electron affinity and the donor
work function, the larger the obtained adhesion energies. This
indicates a significant contribution to the binding between
graphene and ionic mica surfaces resulting from the charge
transfer.

\begin{figure}[!tbp]
\includegraphics[width=0.42\textwidth, angle=0]{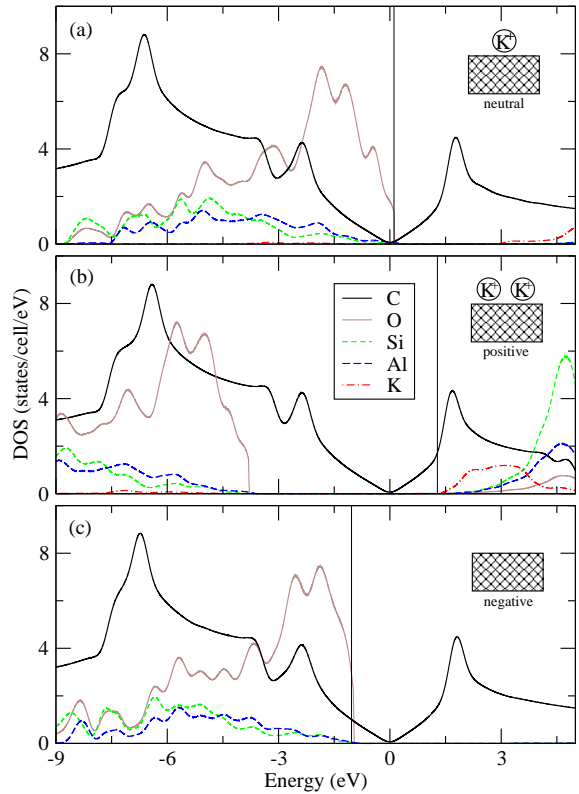}
\caption{(Color online) Projected electronic density of states for graphene deposited on the (a) neutral mica surface, (b) positive mica surface, and (c) negative mica surface (see Sec.~II~B for details).
Oxygen DOS is reduced by a factor of 5 for clarity. Zero energy corresponds to the Dirac point of graphene. The vertical line accentuates the Fermi level.}
\label{mica_dos}
\end{figure}

In order to examine electronic transfer and related properties in
more detail we analyze the density of electronic states. In
Fig.~\ref{mica_dos} we show the projected DOS on different types of atoms in the
supercell for each of the three surfaces with graphene physisorbed: (a)
the neutral surface, (b) the surface with one excess electron, and (c) the surface
with one excess hole. It is noticeable that for all three cases
the typical conical structure of graphene bands in the vicinity of
the Fermi level remains unperturbed. This means that in contrast
to certain metallic surfaces\cite{Khomyakov} and adatoms,
\cite{Chan} as well as to some reactive monoatomic
adsorbates,\cite{imp_wehling} the mica surface cannot break the
strong $sp^2$ network of carbon atoms and thereby is not able to
form a covalent bonding with graphene. Since the main contribution
to the interaction with graphene is defined by the topmost layer
of the substrate, the non covalent nature of the mica-graphene binding is
consistent with other investigations of graphene interaction with
potassium adatoms.\cite{Chan,jin} The fact that the
shape of graphene DOS is not altered in contact with the mica surface
plays a significant role in terms of the conservation of unique
properties of supported graphene. In this respect the mica
surface, having perfect cleavage and atomic smoothness, might be
considered as a substrate for potential graphene-based devices.

For the neutral substrate [Fig.~\ref{mica_dos}(a)] the Fermi level lies above the
valence band formed by 2$p$ electrons of oxygen. In respect to the
conical (Dirac) point, the Fermi level is shifted upward by $\Delta
E_F=$0.1 eV, which implies only a small electron transfer toward
graphene, as expected for inert surfaces. The absence of
hybridization and insignificant charge transfer also indicate a
vdW nature of relatively weak interaction between graphene and
neutral mica surfaces.

For ionic mica surfaces [Figs.~\ref{mica_dos}(a) and \ref{mica_dos}(b)] 
there is a distinct shift of the Fermi-level relative to the Dirac point. 
In the case of positively charged surfaces the electrons are moved toward
graphene and the corresponding Fermi-level shift is $\Delta E_F=$1.3
eV. As a result the $4s$ orbital of the topmost potassium layer
becomes unoccupied. The
opposite situation takes place for the negative surface, where
graphene works as a donor causing the Fermi level shift downward
by $\Delta E_F=$1.0 eV. Despite the electron transfer from
graphene, the oxygen valence band is not completely filled in this
case, indicating the presence of unsaturated oxygen electrons in
the system. As a rule, the charge transfer leads to ionic interactions 
between charged constituents of the system. Therefore, besides the vdW
interaction one can distinguish two different mechanisms resulting
in the binding between graphene and mica, namely, charge transfer
and consequent ionic interaction.

L\"{o}wdin charge analysis\cite{lowdin} shows the
following electron transfers between graphene and mica (in
$e^{-}$/cell): 0.02, 0.91, and $-$0.50, respectively, for neutrally,
positively, and negatively charged mica surfaces. These values are
consistent with previous estimations based on DOS analysis. 
Non equivalence of the charge transfer for differently charged
surfaces allow one to expect an electron-hole asymmetry in graphene
supported on realistic mica. Indeed, there is an 
excess of the electrons transferred from potassium ions toward 
graphene over the hole transfer from potassium-free regions of mica.
On a large scale such asymmetry would produce a shift of the Fermi level 
toward higher energies relative to the Dirac point. In turn, this may
significantly increase the conductivity of mica-supported graphene.

Induced charges in graphene supported on mica might be 
considered as Coulomb impurities and provide a way for charge-carrier scattering.
Recent experiments on potassium-doped graphene demonstrate that
carrier mobility is inversely proportional to the concentration of
potassium ions on the surface.\cite{Chen} 
Moreover, the contribution of potassium ions to the resistivity
is maximal for their homogeneous distribution and
can be strongly suppressed by clusterization.\cite{Cluster1,Cluster2}
In our case this means
that electron mobility should be closely dependent on the particular 
distribution of potassium ions on mica. However, quantitative analysis
of these phenomena requires detailed information about the structure
of the mica surface. We leave this question open for further 
experimental and theoretical studies.

An assumption of a non uniform distribution of potassium atoms on
the mica surface provides a mechanism for height variations of
graphene supported on mica. As can be seen from Fig.~\ref{mica_graphene} 
and Table \ref{mica_table}, the distance from graphene to the topmost
tetrahedral layer of mica is larger when potassium atoms are
present on the surface. For two limiting cases of high potassium
concentration and potassium-free surface the variation of distance
is equal to 1.5 \AA. As follows from the experimental AFM topographic
data,\cite{chun} the upper limit of experimentally observed height
variations corresponds to our variation of the distance between graphene
and the different types of mica surfaces (1.5 \AA) with probability around
99\%. This correspondence between theoretical results
and experimental data allows us to conclude that the irregularity of potassium
distribution on the substrate plays a major role in the formation of the graphene
corrugations on mica. Intrinsic corrugations of graphene\cite{fasolino} are 
expected not to exceed height variations caused by these irregularities.

\section{Conclusion}

We have carried out a first-principles investigation of graphene
supported on a muscovite mica surface using the vdW-DF approach. We
have shown that an assumption of a non uniform distribution of
potassium atoms on the mica substrate may lead to local regions
with an uncompensated charge on the surface. In turn, the presence of
the surface charges significantly affects the adhesion with
graphene. We have found that in the case of the neutral mica surface
the interaction with graphene is mainly of a vdW nature, whereas for
ionic surfaces there are additional contributions arising from the
transfer doping and ionic interaction.

A non uniform distribution of potassium atoms over the surface also provides
the main mechanism for variations of graphene height on mica. Our estimations
show that the obtained theoretical variations are consistent with recent 
experimental data.

Finally, it is important that the
typical shape of a graphene electronic structure remains unchanged
while graphene is deposited on mica. This makes mica a potential
candidate for its use as a substrate for graphene-based
electronics, in spite of the fact that induced charge impurities
may somewhat restrict the unique transport properties of graphene.

\section{Acknowledgments}

We would like to thank Tim Wehling for helpful discussions. The
authors acknowledge support from the Cluster of Excellence
``Nanospintronics'' (Hamburg, Germany), from Stichting voor
Fundamenteel Onderzoek der Materie (FOM, the Netherlands)
and from the Russian scientific programs No. 02.740.11.0217 and 2.1.1/779.

\end{document}